\documentclass[aps,pra,reprint,a4paper,superscriptaddress,numbers,twocolumn,longbibliography,showpacs,showkeys,nobalancelastpage]{revtex4-1}

\usepackage[pdftex]{hyperref,color,graphicx}
\usepackage{amsfonts,amssymb,amsmath}
\usepackage[separate-uncertainty=false]{siunitx}
\usepackage[english]{babel}
\usepackage[utf8]{inputenc}
\usepackage[T1]{fontenc}

\DeclareSIUnit\gauss{G}
\DeclareSIUnit\centimeter{cm}

\hypersetup{
	pdftitle = {Quantum walk of a Bose-Einstein condensate in the Brillouin zone},
	pdfauthor = {Andrea Alberti and Sandro Wimberger},
	colorlinks=true,linkcolor=blue,citecolor=blue,filecolor=blue,urlcolor=blue
}

\newcommand{\figref}[2]{\hyperref[#1]{\getrefnumber{#1}(#2)}}

\addto\captionsenglish{}
\addto\captionsenglish{}

\newcommand{\encapsulateMath}[1]{\raisebox{0pt}[0pt][0pt]{#1}}

\usepackage{xspace}
\usepackage{dcolumn}
\usepackage{bm}
\usepackage{upgreek}
\usepackage{color}
\usepackage{cancel}
\usepackage{braket}
\newcommand{\mb}[1]{{\rm #1}}
\newcommand{\ii}{\;\! \mbox{i} \;\!}

\newcommand{\ic}{\mathrm{i}}

\renewcommand\textemdash{\leavevmode\unskip\kern0.8pt\rule[0.19\baselineskip]{8pt}{0.4pt}\kern1pt\ignorespaces}

\begin{document}

\title{Quantum walk of a Bose-Einstein condensate in the Brillouin zone}

\author{Andrea Alberti}
\email{alberti@iap.uni-bonn.de}
\affiliation{Institut f\"ur Angewandte Physik, Universit\"at Bonn, Wegelerstraße 8, 53115 Bonn, Germany}

\author{Sandro Wimberger}
\email{sandromarcel.wimberger@unipr.it}
\affiliation{Dipartimento di Scienze Matematiche, Fisiche e Informatiche, Universit\`{a} di Parma, Parco Area delle Scienze n.\ 7/a, 43124 Parma, Italy}
\affiliation{INFN, Sezione di Milano Bicocca, Gruppo Collegato di Parma, Parma, Italy}
\affiliation{ITP, Heidelberg University, Philosophenweg 12, 69120 Heidelberg, Germany}

\bibliographystyle{apsrev4-1}

\begin{abstract}
We propose a realistic scheme to implement discrete-time quantum walks in the Brillouin zone (i.e., in quasimomentum space) with a spinor Bose-Einstein condensate. Relying on a static optical lattice to suppress tunneling in real space, the condensate is  displaced in quasimomentum space in discrete steps conditioned upon the internal state of the atoms, while short pulses periodically couple the internal states.
We show that tunable twisted boundary conditions can be implemented in a fully natural way by exploiting the periodicity of the Brillouin zone.
The proposed setup does not suffer from off-resonant scattering of photons and could allow a robust implementation of quantum walks with  several tens of steps at least.  In addition, onsite atom-atom interactions can be used to simulate interactions with infinitely long range in the Brillouin zone.
\end{abstract}%

 
                              \date{\today}

\maketitle

\section{Introduction}%
Quantum walks  form one of the possible bases for universal quantum computing \cite{Childs2009}, for simulating topological phases \cite{Kitagawa:2010,Groh:2016}, for engineering quantum algorithms \cite{Duer2002,kempe2003b}, and have been considered as a possible mechanism explaining the efficient energy transfer in biomolecular clusters \cite{Scholes2012}.  Moreover, the crossover between classical and quantum motion in the presence of decoherence can be optimally studied in quantum walks \cite{Duer2002,Bonn2014,Weiss2015,SW2016} for the reason that the characteristic features of a quantum walk are based on sensitive quantum interference \cite{kempe2003}.

\begin{figure}[tb!]
\begin{center}
\includegraphics[width=\columnwidth]{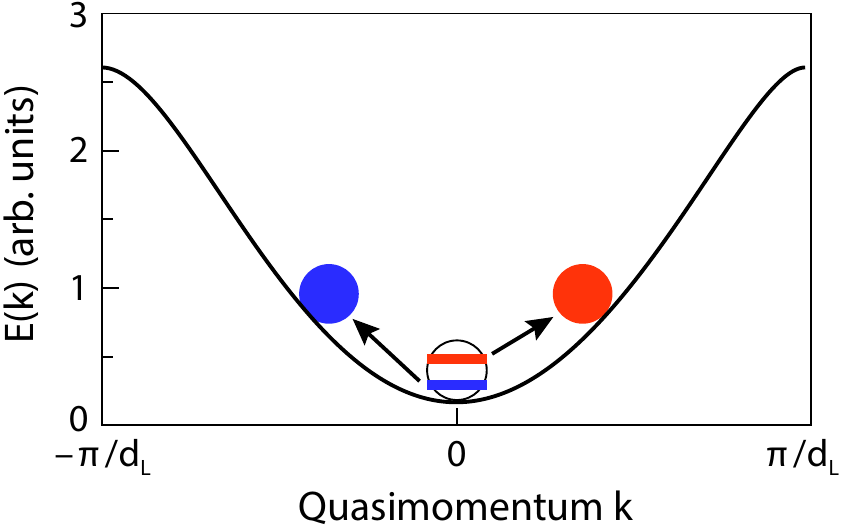}
\caption{\label{fig:1}  Sketch of the proposed experiment for the realization of  quantum walks in the Brillouin zone.
A state-dependent force shifts the atoms to the left or to the right depending on the internal state (sketched by the two arrows). 
The two internal states of the atoms are coherently coupled by short  pulses to realize the coin-toss operation at constant time intervals.
The curve represents the energy dispersion relation $E(k)$ of the populated band.
}
\end{center}
\end{figure}

Inspired by the experimental implementation of quantum walks with trapped neutral atoms \cite{Bonn2009,Preiss2015}, in particular by the experiments carried out  by the Bonn group \cite{Bonn2009,Bonn2015,Bonn2013,Bonn2014,Bonn2015}, we propose a novel realization of discrete-time quantum walks in the reciprocal space of a periodic optical lattice.
Today, it is fairly standard to prepare a Bose-Einstein condensate (BEC) with a vanishing thermal fraction and with a well-defined momentum.
Once loaded into an optical lattice, such a BEC exhibits a very narrow quasimomentum distribution, which can reach down to one percent of the Brillouin zone \cite{Ryu2006}.
In this paper, we propose to delocalize the BEC in a controlled way through discrete steps in the reciprocal space of the lattice. Our scheme realizes a discrete-time quantum walk in quasimomentum space, despite the fact that the motion of atoms in real space is nearly frozen by a deep optical lattice potential:
For that purpose, two internal states of the atoms are employed, which determine the step direction of the BEC in reciprocal space and which are coherently mixed at periodic intervals to realize the so-called coin operation.
As a consequence, the internal states become entangled with the different quasimomentum states of the BEC, which play here the role of the walker's position states.
Like in the Bonn experiments, the internal degree of freedom can be realized through two hyperfine levels of alkaline atoms, forming a pseudo-spin-1/2 system.
In this work, however, we suggest to encode the position of the walker in quasimomentum instead of position space. Figure \ref{fig:1} presents a sketch of the state-dependent shift in a lattice that we envisage. This idea differs from a recent proposal based on a quantum ratchet \cite{SW2016,Gil2016}, where a discrete-time quantum walk is realized in true momentum space by a $\delta$-kicked optical lattice, which leaves quasimomentum unchanged.

Embedding the quantum walk in quasimomentum instead of true momentum space opens new possibilities for investigating interaction-induced effects on the walk \cite{Bonn2012} and novel applications for matter-wave interferometry requiring high spatial densities.
Provided that it is sufficiently deep, the lattice effectively freezes tunneling between lattice sites, thereby ensuring that the original spatial distribution remains invariant for the whole quantum walk evolution.
This implementation of quantum walks opens interesting avenues for future exploration of strongly correlated systems with long-range interactions, which are expected to occur in the reciprocal space (i.e., the effective space for the quantum walk) as a result of localized on-site interactions among ultracold atoms \cite{Gadway:2017}.
Moreover, we expect that our suggestion of quantum walks in the Brillouin zone can be implemented with technology presently available in state-of-the-art BEC apparatuses.

\section{Quantum walks in quasimomentum space}%
\subsection{Physical model without interactions}
We are interested in atoms prepared in a static optical lattice and subject to a state-dependent force $F$, which displaces quasimomentum in opposite directions for the two internal states, as shown in Fig.~\ref{fig:1}. The  motion of an effective two-level atom is then described by the spinor Hamiltonian
\begin{equation}
H  = - \frac{\hbar^2}{2M} \frac{\partial^2}{\partial x^2} 
+ V \sin^2\left(\frac{\pi x}{d_\text{L}}\right)- (x-\bar{x})\hspace{0.3pt}F \hspace{0.3pt}\sigma_z  \,,
\label{eq:Ham}
\end{equation}
where $x$ is the center-of-mass coordinate of the atom, $\bar x$ is a fixed parameter defining the point where the Zeeman potential vanishes, $\sigma_z$ is the Pauli matrix acting on the spinor $\vec{\psi}(x)=\{\psi_1(x),\psi_2(x)\}$, and $M$ is the atomic mass. The optical lattice has a lattice constant $d_\text{L}$ and an amplitude $V$.
We here neglect atom-atom interactions, which will be discussed subsequently in a mean-field approach for weak nonlinearities.

The application of $H$ for a certain time $\tau$ realizes the state-dependent shift operation of a discrete-time quantum walk on a line \cite{kempe2003},
\begin{multline}
	\label{eq:shift_operation}
U_\text{shift} = \exp(-i H \tau/\hbar) \approx\\ \sum_{k}\left(\begin{array}{cc} e^{\ic\hspace{0.3pt}\phi_+(k)}\ket{k+\Delta k}\!\bra{k}&0\\0&e^{\ic\hspace{0.3pt}\phi_-(k)}\ket{k-\Delta k}\!\bra{k}\end{array}\right),
\end{multline}
where $k$ is the quasimomentum of the atom, $\Delta k = F\hspace{0.3pt}\tau/\hbar$ is the step size in the Brillouin zone, and $\phi_\pm(k)$ are Peierls phases, which take into account both dynamical and geometrical phase contributions.
The expression on the right-hand side of Eq.~(\ref{eq:shift_operation}) relies on the assumption that the atoms populate initially only one single band (typically the lowest band), and that Landau-Zener tunneling \cite{Holthaus:2000,peik1997} to higher bands can be neglected.
The latter condition is very well fulfilled for sufficiently deep lattices or small forces
\cite{LandauZenerCondition}.
Note that the operator $U_\text{shift}$, instead of displacing the state in position space by a discrete amount of lattice sites such as in Refs.~\cite{Bonn2009,Bonn2013,Bonn2015}, displaces the quasimomentum state $\ket{k}$ in the Brillouin zone by a fixed amount, which is determined by $\Delta k$, in a direction conditioned upon the internal state.
Using quasimomentum space to implement the discrete-time quantum walk has two advantages: (1) the step size can be arbitrarily tuned by choosing the force $F$ and the duration $\tau$; (2) most importantly,  periodical boundary conditions are naturally realized by quasimomentum $k$ in the Brillouin zone, provided that the latter contains an integer number $n$ of sites, $n \hspace{1pt}\Delta k = 2\pi/d_\text{L}$.

The coin-toss operation couples the two internal states through a Rabi-oscillation dynamics, and is repeatedly applied after a fixed  period $\tau$. 
 It realizes the following beam-splitter unitary transformation acting on $\vec{\psi}$, 
\begin{equation}
U_\text{coin}= \begin{pmatrix} \cos(\alpha/2) & \ic \sin(\alpha/2) \\ \ic \sin(\alpha/2) & \cos(\alpha/2) \end{pmatrix},
\label{eq:matrix}
\end{equation}
where $\alpha$ represents the coin angle \cite{kempe2003}; the symmetric 50-50 beam splitter is realized for $\alpha=\pi/2$, corresponding to the Hadamard quantum walk. Such a unitary transformation is specified in the rotating frame of the drive, where the two hyperfine states of an alkali atom are degenerate in energy, and can be readily implemented with microwave pulses \cite{Gil2003,Bonn2009,Bonn2013,Bonn2015} or with two-photon Raman pulses \cite{Dotsenko:2004,Campbell:2010}. These pulses can be applied very rapidly, on the time scale of microseconds for microwaves, as shown in Refs.~\cite{Bonn2009,Bonn2013}, and in an even shorter time for optical Raman transitions \cite{Campbell:2010}.
In the following, we assume  that the coin pulses act `instantaneously' on the time scale of the evolution of the period $\tau$ and uniformly for all positions $x$ over which the BEC extends, or, alternatively, that the force is switched off while the two internal states are coupled. 
In either case, the time-evolved state after $j$ steps of the quantum walk is equal to \encapsulateMath{$\vec{\psi}_{j}=W\vec{\psi}_{j-1}$}, with $W=U_\text{shift}U_\text{coin}$ being the walk operator of the single step.

\subsection{Peierls phases and twisted boundary conditions}
\begin{figure}[t]
\begin{center}
\includegraphics[width=1.025\columnwidth]{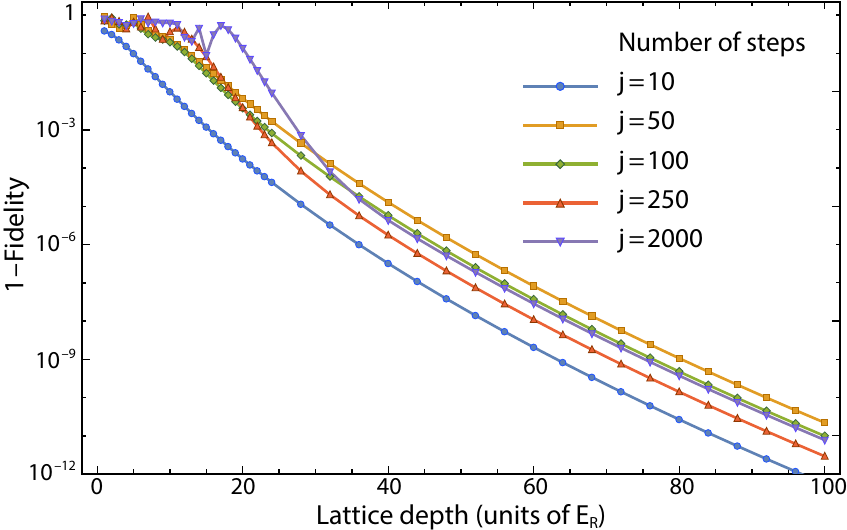}
\caption{\label{fig:2}
(In)fidelity computed as a function of the lattice depth $V$ for a discrete-time quantum walk in reciprocal space with $\num{20}$ sites, for a different number of steps $j$. To compute the fidelity \cite{Fidelity}, a reference walk with vanishing dynamical phases, $\phi^\text{D}_\pm(k)=0$, is considered. Here, the Zeeman force is set to $F=0.2\,E_\text{R}/d_\text{L}$, and periodical boundary conditions are considered with $\bar{x}=0$.
}
\end{center}
\end{figure}

The Peierls phases $\phi_\pm(k)$ are the sum of two contributions, a dynamical $\phi^\text{D}_\pm(k)$ and a geometrical $\phi^\text{G}_\pm(k)$ one:
\begin{eqnarray}
	\label{eq:dynamicalphases}
					\phi^\text{D}_\pm(k) &=& 	\displaystyle \mp\frac{F \hspace{1pt}\bar{x} \hspace{1pt}\tau}{\hbar}\mp \hspace{-2pt} \int_0^{\pm\Delta k} \hspace{-8pt} \text{d}k'\hspace{2pt} E(k+k')/\hbar\,,\\
					\label{eq:geometricalphases}
				\phi^\text{G}_\pm(k) &=&	\displaystyle \hspace{-2pt} \int_0^{\pm\Delta k} \hspace{-8pt} \text{d}k'\hspace{2pt} A(k+ k')\,,
\end{eqnarray}
where $E(k)$ and $A(k)$ are the energy dispersion relation and the Berry-Zak connection \cite{Zak:1989,Xiao:2010} of the occupied band, respectively.
To begin with, we confine ourselves to a subregion of sites strictly within the Brillouin zone $(-\pi/d_\text{L},\pi/d_\text{L}]$, thus excluding that the walker can reach the band edge and wind around the Brillouin zone.
With this constraint, the Peierls phases of the one-dimensional walk operator $W$ can be removed through a gauge transformation when (and only when) the condition 
\begin{equation}
	\label{eq:gaugeawaycondition}
	\phi_+(k)+\phi_-(k+\Delta k)=2\gamma
\end{equation}
is fulfilled \cite{GaugedAwayDynamicalPhases}, with $\gamma$ being a constant term whose only effect is shifting by that amount the global phase of the quantum state after each step.
Peierls phases that can be gauged away have no bearing on observable quantities defined locally in the Brillouin zone such as the probability density (assuming that these phases are maintained static).
The condition in Eq.~(\ref{eq:gaugeawaycondition}) is directly fulfilled by the geometrical phases in Eq.~(\ref{eq:geometricalphases}) and by the first term of Eq.~(\ref{eq:dynamicalphases}), but is not by the second term of Eq.~(\ref{eq:dynamicalphases}). Hence, confined to a subregion of the Brillouin zone, we can neglect the contribution from the geometrical phases as well as that from the dynamical Zeeman phase, but we cannot \emph{a priori} discard the one originating from the energy dispersion $E(k)$.
The energy band's width, however, decreases exponentially with the lattice depth $V$, and the energy band becomes nearly flat for lattice depths of the order of a few tens of recoil energies, $E_{\text{R}}=(\pi\hbar)^2/(2M\hspace{1pt}d_\text{L})$.  
In this limit, the dynamical phases approximately meet the condition in Eq.~(\ref{eq:gaugeawaycondition}) because they reduce to a constant value, which does not depend on $k$ nor on the internal state, and can therefore be neglected.
Figure~\ref{fig:2} shows that neglecting the dynamical phases originating from $E(k)$ is a very good approximation for deep lattices: for example, for $V\hspace{2pt}{>}\hspace{2pt}40\,E_R$, the discrepancy between the evolved quantum state after $j\hspace{2pt}{=}\hspace{2pt}\num{2000}$ steps and that of a discrete-time quantum walk with vanishing Peierls phase results in an infidelity of $1-F\hspace{2pt}{<}\hspace{2pt} \num{e-5}$ \cite{Fidelity}.

Although Peierls phases can be ignored in a subregion of the Brillouin zone when the condition in Eq.~(\ref{eq:gaugeawaycondition}) is met, these can play an important role if the walker is allowed to wind around the Brillouin zone.
In fact, a global gauge canceling all Peierls phases on the entire Brillouin zone  exists only \cite{GaugedAwayDynamicalPhases} when the phase
\begin{equation}
 \label{eq:twistangle}
	\varphi = \sum_k \phi_+(k) \approx -{}n\hspace{1pt} \Delta k \hspace{1pt}\bar{x}+ \int_{-\pi/d_\text{L}}^{\pi/d_\text{L}}\hspace{-3pt} \text{d}k \, A(k)
\end{equation}
is an integer multiple of $2\pi$, with the discrete sum being taken over all $n$ sites in the Brillouin zone. In the the right-hand-side expression, we neglect the dynamical phases originating from energy dispersion since we assume a sufficiently deep lattice, see Fig.~\ref{fig:2}.
The quantity $\varphi$ defines the twist phase of 
twisted boundary conditions~\cite{Xiao:2010},
\begin{equation}
	\braket{k+ 2\pi/d_\text{L}\hspace{0.3pt} | \hspace{0.3pt} \vec{\psi}}=e^{\ic\varphi} \hspace{-1pt}\braket{k \hspace{0.3pt} | \hspace{0.3pt} \vec{\psi}}.
\end{equation}
Twisted boundary
conditions are an important instrument in theory to characterize topological phases of many-body systems \cite{Xiao:2010}, and are generally believed to be difficult to implement in nature \cite{Amico:2005,Grusdt:2014,zoller2016}.
Physically, a quantum walk with twisted boundary condition simulates a particle with charge $Q$ moving on a ring lattice that is threaded by a magnetic flux $\hbar \hspace{1pt}\varphi/Q$.
In this situation, persistent currents are predicted to appear even if no electric field is applied \cite{Zvyagin:2005,Petrescu:2013}, provided that no other relaxation mechanism (e.g., decoherence) is present.

Superlattices such as those demonstrated in Refs.~\cite{Atala:2013,Lohse:2015,Nakajima:2016} allow one to fine tune the twist phase $\varphi$ through the Berry-Zak phase \cite{Zak:1989}, $\phi_\text{Zak}\hspace{2pt}{=}\hspace{2pt}\int \text{d}k \, A(k)$, which is determined by the relative displacement between two commensurate optical standing waves.
For the simple lattice potential considered in Eq.~(\ref{eq:Ham}), the Berry-Zak phase is trivially zero.
However, this is not a limitation since twisted boundary conditions can also be realized by controlling $\bar{x}$ in Eq.~(\ref{eq:Ham}), where in this case $\varphi = 2\pi\hspace{1pt} \bar{x}/d_\text{L}$.
Controlling $\bar{x}$ might be more convenient to be realized with an existing BEC experimental apparatus than controlling the Berry-Zak phase through a superlattice potential.

\subsection{Experimental protocol}%
\begin{figure}[t]
\begin{center}
\includegraphics[width=1.025\columnwidth]{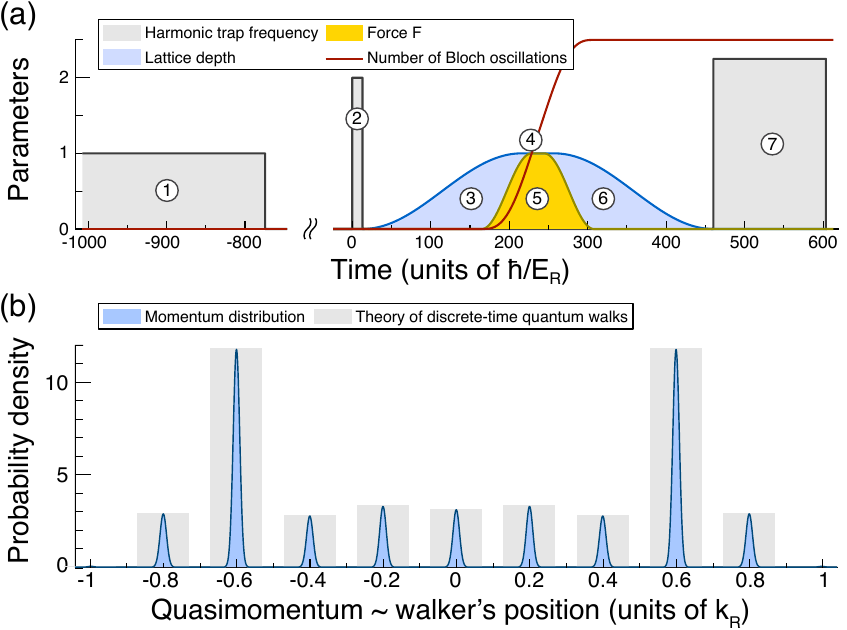}
\caption{\label{fig:3}
(a) Sketch of the experimental protocol to realize discrete-time quantum walks in the Brillouin zone. The curves show in relative units the lattice depth $V$, the force $F$, the harmonic trap frequency $\omega_x$, and the number of Bloch oscillations as a function of time (the initial parameters are provided in the main text). Circled numbers refer to (1) preparation phase in a shallow harmonic trap, (2) free expansion followed by delta-kick cooling, (3) ramp up of lattice and force, (4) pulse preparing internal state, (5) discrete-time quantum walk, (6) ramp down of lattice and force, and (7) Fourier-optics mapping from momentum to position space. (b) (Quasi)momentum distribution normalized to unity after $\num{10}$ steps of the walk in the reciprocal space. The distribution is obtained by integrating numerically the Schrödinger equation, and displayed at the end of the ramp-down phase (6). The underlying gray bars show the theoretical prediction \cite{kempe2003}. In the simulation, coin pulses are applied instantaneously. The temporal evolution of both spatial and momentum distribution is provided in the Supplemental Material~\cite{Figure2Video}.}
\end{center}
\end{figure}

\label{sec:expprotocol}
The full experimental protocol to realize discrete-time quantum walks in the Brillouin zone is illustrated in Fig.~\figref{fig:3}{a} and proceeds as follows: (1) The condensate is first prepared in the ground state of a shallow harmonic trap, in one of the two internal states. (2) It is then allowed to freely expand in a horizontal waveguide potential \cite{McDonald:2013} for a certain time, before the harmonic potential is again turned on for a short transient time; during this transient time the momentum spread is significantly reduced according to the principle of delta-kick cooling \cite{Ammann:1997,McDonald:2013,Muentinga:2013,Kasevich2015}. (3) Subsequently, the condensate is loaded adiabatically into the optical lattice and nearly simultaneously a Zeeman force $F$ is turned on; the duration of this process, in which the lattice potential and the force are ramped up, is chosen such that the quasimomentum sweeps once the Brillouin zone, thus performing a full Bloch oscillation. (4) The internal  state is then prepared in the desired initial superposition of the two internal states  by applying a suitable coin pulse.
(5) The discrete-time quantum walk is performed by $U_\text{shift}$, which shifts the atoms in the Brillouin zone in a direction conditioned upon their internal state , and by applying at periodic time intervals the coin pulse $U_\text{coin}$, which mixes the internal states. (6) To read out the final quasimomentum distribution of the atoms, which is the effective quantum-walk position distribution, the quasimomentum distribution  is mapped onto a momentum distribution by switching off the force and the lattice adiabatically; the switch-off time of the Zeeman force is chosen such that the quasimomentum performs an additional full Bloch oscillation. (7) Finally, to detect the momentum distribution with high resolution, which is required to resolve the small separation $\Delta k$ between two sites of the quantum walk in the Brillouin zone, the atoms are let evolve for a quarter of a period in the shallow harmonic trap \cite{Murthy:2014}. This operation maps in a finite time the momentum distribution onto a position distribution, which can then be readily measured by standard absorption imaging \cite{BDZ2008,Pisa2007,Pisa2008,Pisa2009,Tino2008,Tino2009}.

We note that variations to the proposed scheme are also conceivable: Atoms could be loaded into a vertical optical lattices. The vertical orientation would have the advantage of not requiring a horizontal waveguide potential \cite{McDonald:2013} to support the atoms against gravity. In this case, however, one should keep track of the state-independent shift of quasimomentum caused by the gravity force, which adds on top of the spin-dependent shift produced by the Zeeman force.
Further, in the presence of repulsive atom-atom interactions, which broaden the initial spatial distribution of the BEC \cite{Ryu2006} and correspondingly shrink the initial momentum distribution (see also Sec.~\ref{sec:weak_nonlinearity}), the delta-kick-cooling procedure may even not be required to resolve several lattice sites in quasimomentum space.

To study the behavior of the quantum walker under realistic experimental conditions, we numerically integrate the time-dependent Schrödinger equation bythe split-step finite-difference propagation method \cite{Feit:1982}, assuming the following parameters of the optical lattice: $d_\text{L}= \SI{532}{\nano\meter}$ and $V_0\equiv V/E_\text{R}= 20$, with the recoil energy $E_{\text{R}}=(\hbar \hspace{1pt}k_\text{R})^2/2M$ for $k_{\mb{R}}=\pi/d_\text{L}$. For $^{87}$Rb atoms, these lattice parameters result in a nearly photon negligible scattering rate of $\SI{0.015}{\second^{-1}}$.We adopt rescaled units for quasimomentum $k$ and time $t$  by defining $k_0\equiv  k/k_\text{R}$ and $t_0\equiv E_R\hspace{1pt} t/\hbar$, respectively.
In the rescaled units, the Zeeman force  and the step size in the Brillouin zone read $F_0\equiv Fd_\text{L}/E_\text{R}$ and  $\Delta k_0 \equiv  \Delta k/k_\text{R} = F_0\hspace{1pt}\tau_0/\pi$, respectively, with $\tau_0 \equiv E_\text{R}\hspace{1pt}\tau/\hbar$. For our simulations, we choose $F_0=0.2$, which is in the typical range of previous experiments \cite{Wilk96,peik1997,Pisa2001,Pisa2007,Pisa2008,Pisa2009,Tino2008,Tino2009}, and a longitudinal trap frequency $\omega_x=2\pi \times \SI{10}{\hertz}$, which is applied in the preparation stage.
Moreover, in order to accommodate the entire quantum walk within the first Brillouin zone, we choose the duration of a step (and thus the step size $\Delta k_0$) such that $j\hspace{1pt}\Delta k_0  = 2j/n$ equals one, with $j$ being the number of steps.
Our numerical simulations are based on $^{87}$Rb atoms. However, the choice of the atomic species and other details of the implementation are not crucial, as long as  the parameters remain comparable in the rescaled units, which are marked in the text with the subscript index $0$.
The results of the numerical simulations of the foregoing protocol are shown in Fig.~\figref{fig:3}{b} for $j\hspace{2pt}{=}\hspace{2pt}10$ steps of a Hadamard quantum walk, with the initial internal state chosen in a symmetric superposition.
The numerical results show an excellent agreement with the theory of discrete-time quantum walks \cite{kempe2003}; this is even the case for lattice depths as low as $V_0=2$ (not shown in the figure), provided that dynamical phases $\phi_\pm^\text{D}(k)$ are taken into account in the theoretical model.
For even shallower lattices, the agreement degrades since the condition $V_0\gg \sqrt{32\hspace{1pt}F_0/\pi^2}$ for no Landau-Zener tunneling is not fulfilled \cite{LandauZenerCondition}.

Moreover, the numerical simulations show that the quasimomentum distribution is insensitive to $\bar{x}$  provided that quasimomentum remains confined within the first Brillouin zone, $j\leq n/2$.
Further numerical simulations carried out with $j>n/2$ and $\bar{x}\neq 0$ also exhibit a perfect agreement (not shown in the figure) with the theoretical model of discrete-time quantum walks, corrected to included twisted boundary conditions with twist phase $\varphi=  2\pi\hspace{1pt} \bar{x}/d_\text{L}$.

A crucial condition for future experiments investigating long-range atom-atom interactions beyond the mean-field regime is that the atom density in real space is sufficiently high, and that it remains constant during the entire quantum-walk sequence.
These two conditions are fulfilled in our setup where tunneling of matter waves is frozen for sufficiently deep lattice potentials,  $V_0\gtrsim20$.
The animation provided as Supplemental Material \cite{Figure2Video} shows for  $V_0=20$ that the probability distribution in position space remains unaffected during the quantum walk.

\subsection{Implementation of the Zeeman force}%
\label{sec:imp_force}
A force acting on the atoms can be realized experimentally by gravity \cite{Kasevich1998,Summy1999,Roati2004},  accelerating the standing wave \cite{Wilk96,peik1997,Pisa2001,Pisa2007,Pisa2008,Pisa2009,Tino2008,Tino2009}, state-dependent optical potentials \cite{Deutsch:1998,Jaksch:1999,SW2016,Robens:2017} or  magnetic field gradients \cite{Gust08,Miyake:2013,Aidelsburger:2013}.
Since gravity and lattice accelerations exert equal forces for both spin states, they are not suited for our purpose here.

State-dependent optical potentials do instead discriminate the Zeeman magnetic levels of the atom because they act as an effective magnetic field whose strength depends on the polarization ellipticity \cite{Deutsch:1998}.
Thus, an extra light field superimposed to the optical lattice, which exhibits a polarization gradient from right- to left-handed circular polarization, can be employed to realize the Zeeman force in Eq.~(\ref{eq:Ham}).
Such optical potentials can be switched on and off on the time scale of microseconds, thus allowing one to apply the coin operation while the external degree of freedom does not evolve;
motional excitations produced by the instantaneous switching of the Zeeman force are suppressed when the coin duration  is much shorter than the energy band gap $\approx \pi\hspace{0.3pt}\tau_0/\sqrt{V_0}$, which can be realized using fast coin pulses.
However, state-dependent optical potentials require the laser wavelength to be tuned sufficiently close to an atomic resonance in order to resolve the intrinsic spin-orbit interaction in the excited states of the atom, which is the mechanism underlying their state-dependent action \cite{Steffen:2012,Kennedy:2013};
to mitigate the effects of off-resonant scattering of photons, which is a possible source of decoherence \cite{Bonn2014}, among alkali atoms, heavier ones are more favorable because of their stronger intrinsic spin-orbit coupling.

As an alternative to state-dependent optical potentials, magnetic field gradients can also be used to produce Zeeman forces directed in opposite directions for the two hyperfine states.
The use of magnetic field gradients has the advantage that a only single light field, namely that of the optical lattice, is required to perform quantum walks, and that its wavelength can be very far detuned from any atomic  resonance, resulting in a very little off-resonant scattering of photons.
The downside is that they cannot be switched on and off rapidly.
For our proposal, however, it is sufficient to switch on and off the force at the beginning and the end of the walk,  as shown in Fig.~\ref{fig:3},  provided that the duration of the coin toss is much faster than the duration $\tau$ of the state-dependent shift.
This is certainly doable by choosing a long $\tau$ of the order of $\SI{100}{\micro\second}$ as compared to the duration of a microwave pulse, which can be as short as just a few microseconds \cite{Bonn2009,Bonn2013}.

To use a constant Zeeman force, however, one must also ensure that the pulses act resonantly (i.e., uniformly) for all sites over which the BEC extends despite the spatially dependent detuning caused by the magnetic field gradient. We estimate this detuning across the BEC as follows:
The number $n$ of sites in the Brillouin zone depends on the separation $\Delta k$ between two adjacent sites, $n = 2\pi/(d_\text{L}\hspace{1pt}\Delta k)$;
in order to resolve $n$ sites in the reciprocal space, the BEC must extend in real space, at least, over a number of optical-lattice sites of the same order $n$, meaning that the frequency detuning caused by the Zeeman force across the BEC is of the order of $F\hspace{1pt}n\hspace{1pt}d_\text{L}/\hbar =  2\pi/\tau$. 
Hence, pulses much shorter than $\tau$ do not resolve such detuning and act resonantly over the whole extent of the BEC.

We also note that short pulses, lasting just a few microseconds, could excite higher bands of the lattice since they are spectrally broad and do not resolve the small energy gaps separating the optical lattice bands.
To circumvent this problem, momentum transfer during pulses must be suppressed, either by using microwave radiation or by using copropagating Raman laser beams.
However, suppressing momentum transfer does not suffice alone to prevent excitation of higher bands.
The Zeeman force, in fact, distorts the two optical-lattice potentials trapping the two internal states in a different manner, resulting in a nonvanishing coupling between Bloch eigenstates belonging to different bands \cite{Mintert:2001,Belmechri:2013}.
For sufficiently deep lattices, the effect of the Zeeman distortion is chiefly a state-dependent shift of the trap minimum, which we estimate, by approximating the lattice potential with a harmonic trap, to be about $F_0/(\pi^2\hspace{1pt}U_0)$ in units of $d_\text{L}$.
Using numerical integration of the Schrödinger equation, we confirm the validity of this simple estimate;
moreover, we compute by way of example for $V_0=20$ the overlap integral after adiabatically switching on the Zeeman force between two Bloch eigenstates corresponding to the two different internal states.
We obtain that  the squared modulus of the overlap integral, $|\mathcal{I}|^2$, amounts to $\approx 1-\num{4e-5}$ and $\approx 1-\num{3e-3}$ for $F_0=0.2$ and $F_0=2$, respectively.
Correspondingly, each coin pulse has a probability equal to $1-|\mathcal{I}|^2$ to excite some higher band, limiting the number of coherent steps to about $1/|\mathcal{I}|^2$.

To partially overcome this limitation, one can use a state-dependent optical lattice, which allows  the optical-lattice potentials for the two internal states to be displaced by an amount equal in strength, but opposite in direction to the differential displacement produced by the Zeeman force.
This countermeasure requires choosing  the quantization axis along the lattice direction and setting the polarization of the two counterpropagating laser beams forming the optical lattice potential in a lin-$\theta$-lin configuration \cite{Belmechri:2013}, which can be readily realized using static waveplates.
With reference to the same example of $^{87}$Rb atoms, the Zeeman differential displacement is counteracted by choosing $\theta$ equal to $\SI{12}{\degree}$ and $\SI{66}{\degree}$ for the two cases of the Zeeman force considered above, respectively, resulting in a significantly reduced probability  $\lesssim\num{e-5}$ to excite higher bands for both cases. 

We conclude this section by discussing the requirements on the stability of the Zeeman force that are necessary to prevent decoherence of the discrete-time quantum walk.
Our discussion focuses on the implementation of the Zeeman force by a magnetic-field gradient; analogous considerations may be derived for the implementation by state-dependent optical potentials.

We assume that the magnetic field originates from the superposition of a homogeneous field $B_0$, typically of the order of $\SI{1}{\gauss}$, which defines the quantization axis along the direction of the optical lattice, and a quadrupole field producing a magnetic-field gradient along the same direction.
With reference to the experimental parameters considered in Sec.~\ref{sec:expprotocol}, a force $F_0=0.2$ corresponds to a field gradient $B'$ of about $\SI{10}{\gauss/cm}$ for $^{87}$Rb atoms.

Fluctuations of the gradient strength $B'$ result in fluctuations of the force $F$. These fluctuations cause detrimental variations of the step size, which effectively lead to spin dephasing.
In Appendix~\ref{sec:stepsize} we estimate the decoherence probability per step $p$.
We find that it is related to the relative stability of the field gradient, $p \approx \langle \delta B'^2\rangle/\langle B'\rangle^2$.
Probabilities as low as $10^{-5}$ should be achievable with ordinary laboratory equipment.

Fluctuations of the quantizing magnetic field $B_0$, as well as of stray magnetic fields along the quantization axis, cause stochastic fluctuations of the Zeeman phases in Eq.~(\ref{eq:dynamicalphases}), which lead to dephasing.
In addition, fluctuations of the zero-field position of the quadrupole field also contribute to dephasing since their effect can be represented as an additional fluctuating magnetic field.
We estimate the effect of these fluctuations during shift operations in Appendix~\ref{sec:shiftopdecoh}.
We find that suppression of stray magnetic fields and highly stable current sources may be needed to prevent decoherence.
With state-of-the-art magnetic-field suppression techniques \cite{Dedman:2007,Ruster:2016}, our estimates show that $>100$ coherent quantum-walk steps should be achievable.
In Appendix~\ref{sec:coindecoh}, we also compute the effect of magnetic field fluctuations on the coin operation.
We find that decoherence during the coin operation can be entirely neglected for short pulses.

Note also that slowly fluctuating magnetic fields (i.e., slow drifts) characterized by frequencies much smaller than the inverse of the duration of a quantum walk (typically a few milliseconds) produce no measurable effect on the local properties of a quantum walk on a line \cite{Bonn2014}, because the resulting Zeeman phases, see Eq.~(\ref{eq:dynamicalphases}), can be removed by a suitable gauge transformation \cite{GaugedAwayDynamicalPhases}.
However, this is not the case for quantum walks in quasimomentum space, which take place on a ring geometry instead of a line.
In such a multiply connected lattice, even slow drifts can affect coherences once the walker has wound around the Brillouin zone since the twist phase $\varphi$, see Eq.~(\ref{eq:twistangle}), can vary from one realization of the quantum walk to the next.
Very slow drifts on the time scale of several minutes, or even hours, can be suppressed by calibrating the magnetic-field gradients, for example, by spectroscopic measurements of the atomic hyperfine transitions \cite{Karski:2010}.

\section{Long-wavelength limit and Dirac equation}
A remarkable characteristic of the proposed setup is that it allows one to tune the lattice constant of the effective space of the walk  by simply choosing the step size $\Delta k$.
This can be exploited to simulate `relativistic' continuous-time quantum walks, which are realized in the limiting case when the Zeeman force and the coin pulses act simultaneously.
For sufficiently deep lattices, the dynamics is then described by the one-dimensional Dirac equation,
\begin{equation}
	H_\text{Dirac} \approx  F\hspace{1pt}\sigma_z   (\ii \hbar\hspace{1pt}\partial_k) +   \hbar\hspace{1pt}\Omega_R\hspace{1pt}\sigma_x /2\,,
\end{equation}
where $\Omega_R$ is the Rabi frequency of the coin operation, which in this case is constantly active.
Moreover, by varying $\Delta k$ and the coin angle $\alpha$ one can study the transition of discrete-time quantum walks to the long-wavelength limit of the Dirac equation \cite{Arrighi:2014}.
In the future, it will be interesting to investigate the nonlinear dynamics of the Dirac equation beyond the cases studied in quantum field theory of the Thirring and Gross-Neveu model \cite{Lee:2015}, by including long-range interactions.

\section{Weak nonlinearity}%
\label{sec:weak_nonlinearity}
We study the effect of weak nonlinear interactions on the quantum walk of a condensate in quasimomentum space. In our simulations, we use similar parameters as above in Sec.~\ref{sec:expprotocol}, i.e., $d_\text{L}=\SI{532}{\nano\meter}$, $V_0=10$, and $F_0=\num{0.2}$. Similar values were used as well, for instance, in the experiments reported in Refs.~\cite{Pisa2007,Pisa2008} employing $^{87}$Rb atoms for which the value of the scattering length is $ \approx \SI{5.3}{\nano\meter}$. 

For the small nonlinearities and a quasi-one-dimensional tube geometry considered here, it is sufficient to use a one-dimensional effective theory, which is obtained by rescaling the nonlinear coupling constant as demonstrated in the pioneering paper of Olshanii \cite{O1998}. For simplicity, here we assume that the scattering lengths $a_1=a_2=a_{12}\equiv a_\text{s}$ are equal for all combinations of the two internal states, which is a very good approximation for $^{87}$Rb atoms. As we focus here on the weakly interacting regime, this assumption will not affect our conclusions.  Hence, the dynamics of the spinor condensate can be described by the following time-dependent Gross-Pitaevskii equation simplified to one dimension: 
\begin{multline}
\ii \hbar \frac{\partial }{\partial t}\vec{\psi} (x,t) =
\left[-\frac{\hbar^2}{2M} \frac{\partial^2}{\partial x^2} + V \sin^2\left(\frac{\pi x}{d_\text{L}}\right)
 -F\hspace{0.3pt}x\hspace{0.3pt}\sigma_z\hspace{2pt}+ \right.
 \\
\hspace{-6pt} \left.
  \frac{1}{2}M \omega_x^2 x^2  + g_\text{1D}  \left( \left| \psi_1(x,t) \right|^2 + \left| \psi_2(x,t) \right|^2 \right)
\right] \vec{\psi}(x,t) \,,\hspace{-5pt}
\label{eq:GP1d}
\end{multline}
where we assume cylindrical symmetry as realized, e.g., in Ref.~\cite{Esslinger2003}.
The nonlinearity parameter $g_\text{1D}=2\hspace{1pt}\hbar \hspace{1pt}a_\text{s}\hspace{1pt} \omega_\text{r}\hspace{0.2pt}N$ accounts for the interaction-induced nonlinearity, which depends on the number of atoms $N$ and on the frequency $\omega_\text{r}$ of the radial trap. This expression is valid for $\hbar/(M a_\text{s}^2)\gg\omega_\text{r}\gg \omega_x$. In our simulations, we chose $\omega_\text{r}=2\pi\times \SI{100}{\hertz}$, $\omega_x=2\pi \times \SI{2.5}{\hertz}$. The harmonic confinement in both longitudinal and radial directions can be maintained during the walk evolution without any noticeable effect for a walk of $j\hspace{2pt}{=}\hspace{2pt}\num{10}$ steps, which we consider here. The Gross-Pitaevskii in Eq.~(\ref{eq:GP1d}) is numerically integrated using the finite-difference propagation method, adapted to include a predictor-corrector estimator to reliably evaluate the nonlinear interaction term \cite{Pisa2005,Pisa2005b}. In the simulation, the coin tosses are assumed to be instantaneous or, at least, much faster than $\tau$. 

\begin{figure}[t]
\begin{center}\includegraphics[width=\columnwidth]{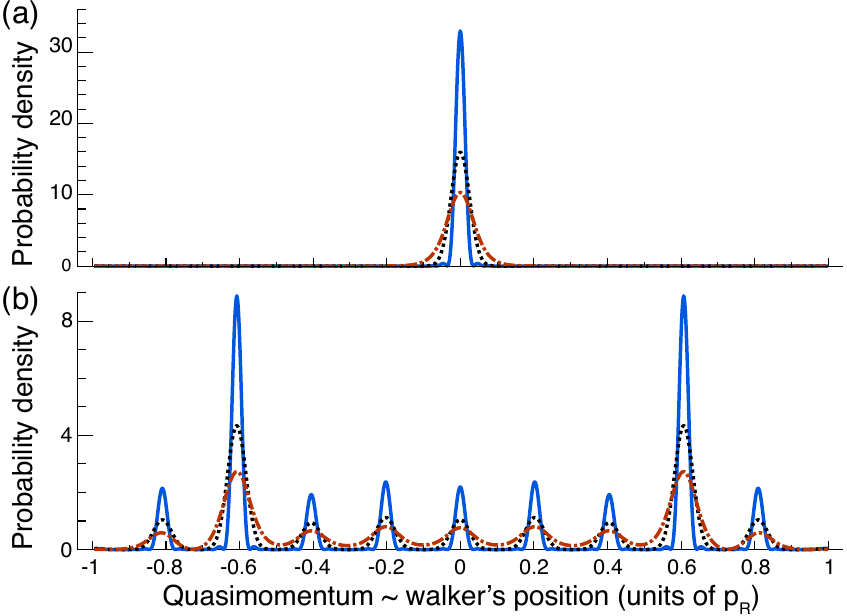}
\caption{\label{fig:4}
Quasimomentum distributions  normalized to unity.
(a) Bose-Einstein condensate prepared initially in the ground state of an optical lattice, i.e., at the minimum of the band structure, where the
black dotted line is computed without interactions, whereas the blue solid and  red dash-dotted lines are obtained for repulsive and attractive nonlinearity, respectively.
(b) Quantum-walk distribution obtained after ten steps  for linear evolution, repulsive and attractive interactions integrating Eq.~\eqref{eq:GP1d}. All parameters are given in the text. 
}
\end{center}
\end{figure}

The results of the numerical simulations of a discrete-time quantum walk with weak nonlinearities are presented in Fig.~\ref{fig:4}. The simulated quasimomentum distribution shows that tens of steps of a quantum walk should be well within the reach of present BEC experimental apparatuses. 
These results are obtained for relatively small nonlinearities, with $a_\text{S}=\pm\SI{5.3}{\nano\meter}$ and about $\num{100}$ atoms per tube for the positive and 10 atoms for the negative value of the scattering length, i.e., for  repulsive and attractive interactions, respectively; stronger nonlinearities would require other integration methods suited for strongly correlated quantum systems.

Indeed, our simulations show that the main effect of weak nonlinearities  is merely influencing the width of the initial momentum distribution.
Interestingly, the repulsive nonlinearity enhances the effective spatial resolution of the quantum walk since the momentum peaks tend to be narrower (as opposed to the distribution in real space, which is broadened). As seen in Fig.~\ref{fig:4}, this effect can be used to prepare narrower initial peaks and thereby to host a higher number of sites in the Brillouin zone. In contrast, even very weak attractive nonlinearities produce the opposite effect of decreasing the visibility of the peaks by broadening them substantially. This is related to the tendency to collapse in real space, resulting in an instability of the condensate for attractive interactions \cite{Cornell2001,Pisa2005b}.

\section{Conclusions and outlook}%
Our work shows that  atoms can be frozen in real space using a sufficiently deep optical lattice, and yet can perform a discrete-time quantum walk in reciprocal space.
A robust protocol to realize  quantum walks in quasimomentum space is provided, including  estimates of parameters and the numerical simulations based on realistic experimental conditions.
Employing an optical lattice to freeze the atoms' motion in real space holds promise to overcome present decoherence limitations of momentum-space experiments with nontrapped particles, which suffer from spatial separation of wave packets possessing different momentum components~\cite{Meier:2016}.

The present work only focuses on a weakly interacting BEC, which is described by a one-dimensional single-mode mean-field approach.
In this regime, we observe that the different quasimomentum peaks remain well separated and their relative populations are unaffected by interactions.

An interesting open question concerns the effect of stronger atom-atom interactions. Contact interactions in real space are expected to produce infinitely long-range effects in quasimomentum space. This regime completely differs from that of strongly correlated quantum walks in real space \cite{Bonn2012,Schreiber:2012,Preiss2015}. In momentum space, the long-range interactions may not just dephase the ideal walk considered here, but might lead to novel topologically ordered states and to other interesting mechanisms, for example, to bias or to steer the evolution.

As suggested in Ref.~\cite{Gadway:2017}, effective finite range interactions may emerge as a result of quantum statistics (i.e., exchange interaction) even when the scattering lengths are isotropic, as is the case of $^{87}$Rb atoms.
Such a study of interaction effects calls for a more complex theory allowing one to capture strong correlations in the system, possibly including more degrees of freedom such as spin mixtures.

In addition, we suggest the possibility to  realize discrete-time quantum walks with more than two internal states.
This could be achieved with the proposed setup by exploiting a larger manyfold of Zeeman states, where the state-dependent shift operation displaces the atoms by a discrete number of sites that is proportional to the Zeeman quantum number.

\begin{appendix}
\section{Decoherence analysis}
In this appendix, we estimate the decoherence effects produced by a fluctuating Zeeman force.
We focus here on the implementation of the Zeeman force by a magnetic field gradient;
however, similar results can be obtained for the case in which the Zeeman force is implemented by state-dependent optical potentials, see Sec.~\ref{sec:expprotocol}.

\subsection{Fluctuating step size}
\label{sec:stepsize}
Fluctuations of the Zeeman force $F$, here denoted by $\delta F$, make the step size become a stochastic quantity,
\begin{equation}
\int_0^{\tau} \hspace{-1pt}\mathrm{d}t\,[F+\delta F(t)]/\hbar = 	\Delta k +\delta  k,
\end{equation}
where $\delta k$ is the fluctuating component characterized by a zero mean value and by a variance
\begin{equation}
		\label{eq:variance_step_size}
	\langle\delta k^2\rangle=\int_{\omega_\text{c}}^{\infty}\hspace{-3pt}\mathrm{d}\omega\, \pi f(\omega) S_{F}(\omega)\hspace{1pt}\tau/\hbar^2\,.
\end{equation}
In this expression, $\omega_c$ is a low-frequency cut-off amounting approximately to the inverse of the overall duration of a quantum walk experiment (or, more generally, the inverse of the time between two subsequent calibration operations of the field gradient), $S_{F}(\omega)$ is the noise spectral density normalized such that \encapsulateMath{$\int_{\omega_c}^\infty \hspace{-3pt} \mathrm{d}\omega\, S_{F}(\omega)=\langle \delta F^2\rangle$} is equal to the variance of $\delta F$, and $f(\omega)=\tau\operatorname{sinc}^2(\omega \hspace{0.3pt} \tau/2)/\pi$ is the window function  normalized to unity, $\int_0^\infty \hspace{-3pt} \mathrm{d}\omega\,f(\omega)=1$.
The window function makes the step size insensitive to force fluctuations $\delta F$ occurring at frequencies much higher than $1/\tau$, since these fluctuations produce a vanishing effect on average.

To estimate the decoherence effect of the fluctuating step size, we assume that each quasimomentum peak of the quantum walk in reciprocal space is characterized by a Gaussian wave function with a width of $\sigma_k \equiv 2\pi/(\beta \hspace{0.5pt}n)$, where $n$ is the number of sites in reciprocal space and $\beta>1$ is a small number denoting how well a single site is resolved.
Thus, the overlap between the wave function shifted by $\delta k$ in quasimomentum and the original wave function is reduced to less than one, with its modulus squared being equal to $\exp[-\delta k^2/(4\sigma_k^2)]\approx 1-\delta k^2/(4\sigma_k^2)$.
This decoherence process can be effectively modeled as spin dephasing, with a decoherence probability per step equal to $p\approx\langle\delta k^2\rangle/(4\sigma_k^2) + \mathcal{O}[(\langle\delta k^2\rangle/\sigma_k^2)^4]$ in the limit of fluctuations with a small rms deviation.

For a magnetic-field gradient producing the force $F$, we expect that fluctuations of the force predominantly occur at frequencies smaller than $1/\tau$, with $\tau$ being of the order of $\SI{100}{\micro\second}$, see Sec.~\ref{sec:expprotocol}.
We can thus replace $f(\omega)$ in Eq.~(\ref{eq:variance_step_size}) with its dc value, $f(0)>f(\omega)$, to obtain an upper bound for the decoherence per step, $p \lesssim (\beta/2)^2\, \langle\delta F^2\rangle /\langle F\rangle^2 =(\beta/2)^2 \langle\delta B'^2\rangle /\langle B'\rangle^2$, where $B'$ denotes the strength of the magnetic-field gradient. 
From this result, we obtain that a decoherence probability per step of $p=\num{e-5}$ corresponds to relative fluctuations of the field gradient of about $\sqrt{\langle\delta B'^2\rangle} /\langle B'\rangle \approx \SI{0.4}{\percent}$.
Because such a gradient stability can be achieved with ordinary laboratory equipment, it can be concluded that decoherence by magnetic-field gradient fluctuations can be neglected in experiments.

\subsection{Decoherence during shift operations}
\label{sec:shiftopdecoh}
At the location of atoms, a nonvanishing magnetic field is present to define the quantization axis.
During the state-dependent shift operation, fluctuations of the magnetic field along the quantization axis result in the accumulation of a stochastic relative phase between the two internal states, which causes decoherence by spin dephasing \cite{Bonn2014}.
The overall relative phase, which is accumulated in a shift operation, amounts to
\begin{equation}
-\hspace{-2pt}\int_0^{\tau} \hspace{-1pt}\mathrm{d}t\,2\hspace{0.3pt}F(t)\hspace{0.3pt}\bar{x}(t)/\hbar =\bar\phi + \delta\bar{\phi},
\end{equation}
where $\bar\phi \equiv -2\hspace{1pt} \langle F\hspace{1pt}\bar{x}\rangle\hspace{1pt} \tau/\hbar$ is the Zeeman contribution to the dynamical phase difference $\phi^\text{D}_+(k)-\phi^\text{D}_-(k)$, see Eq.~(\ref{eq:dynamicalphases}).

We assume that the magnetic field along the quantization axis takes the form of $B(x)=(x-\xi)B'+B_0$.
Such a field produces the Zeeman potential introduced in Eq.~(\ref{eq:Ham}), $-(x-
\bar{x})F\sigma_z=(x-{\bar{x}})\hspace{0.3pt}B'\mu_B\hspace{0.3pt}m_F\hspace{0.3pt}g_F\hspace{0.3pt}\sigma_z$, with $\bar{x} = \xi - B_0/B' + \Xi$. Here, $B_0$ is the strength of a homogeneous magnetic field, $B'$ is the strength of the magnetic field gradient, $\xi$ is the zero-field position of field gradient, $\Xi$ is a constant determined by the frequency of the rotating frame, $\mu_B$ is the Bohr magneton, and $\pm g_F \hspace{0.3pt} m_F$ is the product between the Landé factor and the magnetic quantum number for the two internal states.
The parameter $\Xi$ is chosen such that $\bar{x}$ lies close to the center of the BEC to allow the coin pulses to act resonantly at all sites over which the condensate extends, see Sec.~\ref{sec:imp_force}.

The magnetic field gradient is generated by a quadrupole magnetic field produced by, e.g., a pair of anti Helmholtz coils or by a permanent quadrupole magnet.
We assume that the zero-field position $\xi$ of such a quadrupole field lies, by construction, at some point close to the center of the BEC.
In reality this point is not fixed, but can fluctuate with respect to the optical lattice where the atoms are trapped. These fluctuations can be accounted for by including them in $\delta\bar\phi$, as shown below.

The homogeneous magnetic field $B_0$, which is used to define the quantization axis, is generated by a pair of Helmholtz coils and is assumed to be of the order of $\SI{1}{\gauss}$, about two orders of magnitude higher than the field strength $B'\xi$ produced by the field gradient.

The accumulated relative phase $\delta\bar\phi$ is a stochastic quantity with variance 
\begin{equation}
		\label{eq:variance_phase}
\langle\delta \phi^2\rangle=
\int_{\omega_\text{c}}^{\infty}\hspace{-3pt}\mathrm{d}\omega\, \pi f(\omega)
	S_{B'\bar{x}}(\omega) \hspace{1pt} \tau \hspace{1pt} (2\mu_B\hspace{0.3pt}m_F\hspace{0.3pt}g_F/\hbar)^2\,
\end{equation}
where $S_{B'\bar{x}}(\omega)$ is the noise spectral density of $B'\bar{x}$.
Such a spectral density is the sum of two contributions, $S_{B'\bar{x}}(\omega)=S_{B_0}(\omega) + S_{B'\xi}(\omega)$, with $S_{B_0}(\omega)$ originating from fluctuations of the quantizing magnetic field $B_0$, with normalization \encapsulateMath{$\int_{\omega_c}^\infty \hspace{-3pt} \mathrm{d}\omega\, S_{B_0}(\omega) = \langle \delta B_0^2\rangle$}, and $S_{B'\xi}(\omega)$ originating from fluctuations of $B'\xi$, with related normalization. 

Assuming a relative stability at around $10^{-5}$ for the current sources generating the quantizing magnetic field, we infer a rms deviation of the magnetic field at around \encapsulateMath{$\sqrt{\langle \delta B_0^2 \rangle}\approx 10^{-5}\langle B_0\rangle \approx  \SI{10}{\micro\gauss}$}.
Note that, under typical laboratory conditions, $S_{B_0}(\omega)$ also contains a contribution from fluctuating stray magnetic fields oriented along the quantization axis, which can be as large as the contribution from fluctuations of the current sources \cite{Ruster:2016}.
By suppressing stray magnetic fields using a magnetic field shielding and by generating the quantizing magnetic field with permanent magnets, a coherence time of a Zeeman qubit (without spin echo) of $\SI{300}{\milli\second}$ has been reported \cite{Ruster:2016}, which corresponds to \encapsulateMath{$\sqrt{\langle \delta B_0^2 \rangle}\approx \SI{0.1}{\micro\gauss}$}.

Further, $S_{B'\xi}(\omega) = B'^2 S_{\xi}(\omega) + \xi^2  S_{B'}(\omega)$ is the sum of two contributions originating from fluctuations of $\xi$ and $B'$, respectively.
Because the average field $\langle B'\xi \rangle$ is two orders of magnitude smaller than $\langle B_0\rangle$ (see above) we obtain $\xi^2 S_{B'}(\omega)\approx 10^{-4} S_{B_0}(\omega)$ under the assumption that the current sources generating the dipole and quadrupole fields have the same relative stability.

Very low-frequency fluctuations (i.e., slow drifts) of $\xi$ can be suppressed by calibrating the magnetic field gradient, for example, by means of spectroscopic measurements of the atomic hyperfine transitions.
Low frequency fluctuations can be suppressed by actively stabilizing \cite{Robens:2016c} the position of the optical standing wave producing the lattice with respect to the position of the coils producing the field gradient;
measurements \cite{Robens:2016c} show that the residual fluctuations of $\xi$ have a rms deviation \encapsulateMath{$\sqrt{\langle \delta\xi^2\rangle} \lesssim \SI{100}{\pico\meter}$}, where $\langle \delta\xi^2\rangle$ is experimentally obtained by integrating $S_\xi(\omega)$ over a 10-$\si{\mega\hertz}$ bandwidth.
Hence, we expect that fluctuations of $\xi$ result in a rms deviation of $B'\sqrt{\langle \delta\xi^2\rangle} \lesssim \SI{0.1}{\micro\gauss}$, where we assume $B'=\SI{10}{\gauss/\centimeter}$, see Sec.~\ref{sec:imp_force}.

Thus, if the noise spectrum $S_{B_0}(\omega)$ is concentrated at frequencies smaller than $1/\tau$, we obtain an upper bound of the phase deviation, $\langle\delta \phi^2\rangle \lesssim (2\mu_B\hspace{0.3pt}m_F\hspace{0.3pt}g_F\hspace{0.3pt}\tau/\hbar)^2 \langle \delta(B'\bar{x})^2\rangle$, by replacing  $f(\omega)$ with its dc value in Eq.~(\ref{eq:variance_phase}).
Note that for even longer $\tau$ exceeding the correlation time of $B'\bar{x}$, the window function $f(\omega)$ modifies the scaling behavior of $\langle\delta \phi^2\rangle$ from $\langle\delta \phi^2\rangle\,{\propto}\, \tau^2$ to $\langle\delta \phi^2\rangle\,{\propto}\, \tau$.
Coherences \cite{Bonn2014,Ruster:2016} are suppressed by a factor $C=\langle \exp(i\hspace{1pt}\delta\bar{\phi})\rangle$, which can be computed as $\exp(-\langle\delta \bar{\phi}^2\rangle/2)$ provided that the relative phase $\delta \bar{\phi}$ is Gaussian distributed [which is the case when the noise spectral power $S_{\bar{x}}(\omega)$ is not concentrated in a few single-frequency peaks].
Hence, we estimate that the number of coherent steps is of the order of
\begin{equation}
	\frac{\hbar}{2\mu_B\hspace{0.3pt}m_F\hspace{0.3pt}g_F\hspace{0.3pt}\tau\hspace{0.3pt}\sqrt{\langle \delta(B'\bar{x})^2\rangle}} \approx 10^3
\end{equation}
based on the assumptions of
\encapsulateMath{$\sqrt{\langle \delta(B'\bar{x})^2\rangle}\approx \SI{1}{\micro\gauss}$} and $\tau=\SI{100}{\micro\second}$, see Sec.~\ref{sec:expprotocol}.

\subsection{Decoherence during coin operations}
\label{sec:coindecoh}
Errors can also occur during the coin operation because of fluctuations of magnetic fields.
To evaluate coin errors, we have computed the evolution of the internal state for a Hadamard pulse in the presence of a fluctuating stray magnetic $B$ field which is characterized by a noise spectral density $S_B(\omega)$.
The calculation is carried out analytically up to the second order in the amplitude of the fluctuating magnetic field [i.e., first order in $S_B(\omega)$].
By solving the evolution for an arbitrary initial density matrix operator, we can determine the quantum process matrix and the operator-sum representation of the decohered coin operation in terms of the Kraus operators.
The details of the calculation will be reported elsewhere. Here, we summarize the main result.

From the computed expression of the quantum process matrix, we obtain the process fidelity \cite{Gilchrist:2005},
\begin{eqnarray}
	F_\text{pro}^2&=&1- 2\left(\frac{\mu_B\hspace{0.3pt}m_F\hspace{0.3pt}g_F}{\hbar\hspace{0.3pt}\Omega_R}\right)^2\hspace{-3pt}\int_{\omega_C}^\infty\hspace{-1pt}
	\mathrm{d}\omega\, \hspace{1pt}S_B(\omega)\,
	g(\omega/\Omega_R)\hspace{0.5pt},\hspace{0.5mm}\text{with}\nonumber\\
	 g(r)&=&\frac{1+r^2-2\hspace{1pt}r\sin(\pi r/2)}{(1-r^2)^2}\,
	\end{eqnarray}
where the $r=\omega/\Omega_R$ is the ratio between frequency of the fluctuating magnetic field and the Rabi frequency $\Omega_R$ associated with the Hadamard operation.
Note that $g(\omega/\Omega_R)$, as a function of $\omega$, has a width of about $\Omega_R$, its maximum at zero frequency, and its integral $\int_0^\infty\hspace{-3pt} \mathrm{d} r \, g(r) = \pi^2/4$.
The process fidelity is related to the average fidelity $F_\text{ave}^2 = (1+d\,F_\text{pro}^2)/(d+1)$, where $d=2$ and the average takes place over all pure input states.
Moreover, the process fidelity measures the distance between the physical (decohered) process and the ideal unitary transformation, providing us with an upper bound \cite{Gilchrist:2005} on the probability of an error by the coin operation, $p=1-F_\text{pro}^2$.
Assuming \encapsulateMath{$\sqrt{\langle \delta B^2\rangle}=\SI{1}{\micro\gauss}$}, see Sec.~\ref{sec:shiftopdecoh}, and $\Omega_R=2\pi\times \SI{200}{\kilo\hertz}$, we obtain that the error probability $p\lesssim 10^{-10}$ is vanishingly small.

\end{appendix}

\begin{acknowledgments}
A.A. acknowledges financial support from the SFB/TR 185 OSCAR and from the ERC grant DQSIM. He is thankful to Wolfgang Alt for insightful discussions.
\end{acknowledgments}

\bibliographystyle{apsrev4-1}
\bibliography{BF}

\begin{thebibliography}{79}
\bibitem [{Gau()}]{GaugedAwayDynamicalPhases}%
  \BibitemOpen
  \bibinfo {note} {Peierls phases can be ``gauged'' away when the phase
  integrated along any closed path vanishes.}\BibitemShut {Stop}%
\bibitem [{Fid()}]{Fidelity}%
  \BibitemOpen
  \bibinfo {note} {The fidelity $F$ between two pure quantum states is
  evaluated as the modulus of their overlap integral.}\BibitemShut {Stop}%
\end{thebibliography}

\end{document}